# Spin-squeezing-enhanced dual-species atom interferometric accelerometer employing large momentum transfer for precision test of the equivalence principle

Jinyang Li[1], Gregório R. M. da Silva[1], Schuyler Kain[1], Jason Bonacum[3], David D. Smith[4], Timothy Kovachy[1], and Selim M. Shahriar[2,3]

[1] Department of Physics and Astronomy, Northwestern University, Evanston, IL 60208, USA
[2] Department of Electrical and Computer Engineering, Northwestern University, Evanston, IL 60208, USA
[3] Digital Optics Technologies, Rolling Meadows, IL 60008, USA
[4] NASA Marshall Space Flight Center, Space Systems Department, ES23, Huntsville, Alabama 35812, USA

## Abstract

We theoretically investigate the feasibility of applying spin squeezing to a light pulse atom interferometer in the presence of large momentum transfer using off-resonant Raman transitions, in order to enhance the sensitivity of accelerometry close to the Heisenberg limit. We also show how to implement this scheme in a dual-species atom interferometer for precision test of the equivalence principle by measuring the Eötvös parameter, and identify the spin squeezing protocol that is best suited for such an experiment. For a space borne platform in low earth orbit, such a scheme may eventually enable the measurement of the Eötvös parameter with a sensitivity of the order of $10^{-20}$.

## 1. Introduction

Precision test of the equivalence principle (EP), which lies at the heart of general relativity (GR), is one of the grand challenges in fundamental physics. The EP is characterized by the Eötvös parameter, $\eta$, defined as the ratio of the differential acceleration to the mean acceleration experienced by two objects with different inertial masses under free fall in a gravitational field. GR has been immensely successful in predicting many phenomena. However, there are also reasons to believe that GR is incomplete. One indication of this is that GR cannot be reconciled with the standard model (SM). Another indication is that GR cannot explain the accelerated expansion of the universe. EP violation was found by the Committee on the Physics of the Universe to be relevant to four of the eleven questions that are the focus of the landmark study commissioned by the National Research Council [1]. Search for EP violation with ever increasing precision thus attempts to answer these four questions: Did Einstein have the last word on gravity? Are there additional space-time dimensions? What is the nature of dark energy? Is a new theory of matter and light needed at the highest energies?

Fundamentally, a violation of the EP arises when the SM is augmented by a new field with gravitational-strength coupling [2]. Since this field does not couple universally to SM fields, it causes different rates of fall. Such fields are explored in many attempts to account for the accelerated expansion of the universe, and to unify gravity with the other fundamental interactions. Proposals with concrete predictions for EP violation include Lorentz-violating fields introduced as part of the SM extension [3,4], cosmological models where a new field couples indirectly to SM fields via interactions with dark matter [5], chameleon fields with density dependent masses [6,7], and modified-gravity theories in which the new field is related to spacetime curvature [8].

The most precise test of the EP has been carried out under the satellite-borne MICROSCOPE experiment employing classical accelerometers [9,10]. The final result from this experiment [10] constrains the value of $\eta$ to $\sim 1.5 \times 10^{-15}$. Efforts are underway to reach a sensitivity of $10^{-18}$ under the STEP experiment [11]. A thorough analysis of the justification for exploring EP violation at a level of $10^{-18}$ or smaller can be found in Ref. [2]. Here, we propose the use of space-borne dual species atomic interferometry, augmented by spin-squeezing and large momentum transfer, that may make it possible to reach a sensitivity of $\eta < 10^{-20}$ for experimentally accessible parameters.

The conventional dual-species light pulse atom interferometer offers the potential to test the equivalence principle [12, 13, 14, 15, 16, 17, 18, 19, 20, 21] and to search for the wavelike dark matter [22, 23] with high precision. This approach has constrained the value of $\eta$ to $1.5 \times 10^{-12}$ using a terrestrial experiment [13]. It has been estimated that a space-borne version of a similar experiment, employing five parallel dual species atom interferometers employing Bose-Einstein condensates, can reach a sensitivity to measure $\eta$ as small as $10^{-17}$ in 18 months [24]. The technique of large momentum transfer (LMT) [25, 26, 27, 28, 29, 30, 31, 32, 33, 34] has been used for enhancing the sensitivity of a dual-species atom interferometer [12, 13]. However, another technique that can in principle enhance the sensitivity of the atom interferometer, namely spin squeezing [35, 36, 37, 38, 39], has not yet been applied to a dual-species atom interferometer. Furthermore, the combination of LMT and spin squeezing has not been investigated even for the single-species atom interferometer. It should also be noted that for atom interferometry spin squeezing has only been used for suppressing the quantum projection noise [40, 41] which reduces the resistance to detection noise. Alternative approaches of spin squeezing that produces phase magnification [42, 43] to increase the resistance to detection noise has not been explored in the context of atomic interferometry.

In recent years, we have been investigating, separately, the use of LMT in atom interferometry [25] as well as the application of various protocols using cavity-induced one-axis-twist squeezing (OATS) for enhancing the sensitivity of atomic interferometers and atomic clocks [42, 43, 44].

The approach for realizing OATS that has been studied most extensively, theoretically as well as experimentally, is the one based on cavity-mediated interactions [36, 37, 38, 39]. In what follows, we consider, for specificity, only this approach for realizing the OATS scheme. The Hamiltonian of OATS can be expressed as $\hbar\chi S_z^2$, where $S_z$ is the z-component of collective spin operator, and $\chi$ is a characteristic frequency representing the squeezing process. The corresponding propagator can be expressed as $e^{-i\mu S_z^2}$, where $\mu$ is the squeezing parameter, defined as $\chi$ times the interaction time. Arguably the most promising approach for applying OATS to atomic interferometry makes use of the so-called generalized echo squeezing protocol (GESP) [42], which is predicted to yield a sensitivity of the Heisenberg limit within a factor of $\sqrt{2}$. The GESP has two versions. Their behavior does not differ much until $\mu$ approaches $\pi/2$. As $\mu$ approaches $\pi/2$, one version is optimized for even $N$, where $N$ is the number of atoms, and is thus denoted as GESP-e, while the other version is optimized for odd $N$, denoted as GESP-o. Other protocols of interest with similar degree of enhancement in sensitivity include the Schrödinger cat state protocol (SCSP) [44, 43], and the conventional echo squeezing protocol (CESP) [45, 46]. The SCSP also has two versions, namely SCSP-e and SCSP-o, where SCSP-e is optimized for even $N$, and SCSP-o is optimized for odd $N$. Both versions of the GESP becomes identical to the corresponding version of the SCSP for $\mu = \pi/2$. Each of the squeezing protocols mentioned above employs the squeezing operation and the inverse of the squeezing operation.

To apply OATS to the atom interferometer, we first need to determine whether the Raman pulse [47, 48] or the Bragg pulse [31, 34] should be used for the interferometer. In principle, two momentum states with the same internal state coupled by a Bragg transition can also be squeezed [49]. However, squeezing the momentum states using the OATS process may not work well for alkali atoms, mainly because the linewidths of optical transitions (~6 MHz, for example, in Rb), relevant for the three-level system used in OATS, are much larger than the energy difference (~15 kHz for Rb) between the two momentum states coupled by a Bragg transition. In contrast, the two ground states involved in the Raman transition is 3.0 GHz for $^{85}$Rb and 6.8 GHz for $^{87}$Rb, which are much larger than the linewidths of the optical transitions. Therefore, in this paper, we only consider atom interferometers employing Raman pulses.

In Refs. [42, 44, 43], we proposed the SCSP and GESP versions of the squeezing protocols mentioned above, adapted to the atom interferometer. These versions are based on the conventional light pulse atom interferometer that uses the sequence of $\pi/2$, $\pi$, $\pi/2$ counter-propagating Raman pulses. However, use of this sequence for the squeezing protocols lead to a practical problem. The distance between the two states of the atoms produced by the first $\pi/2$ pulse will keep increasing during the squeezing process, which imposes significant constraints on the squeezing process, especially for protocols that requires a relatively long atom-cavity interaction time. Therefore, in this paper we adopt a different scheme of atom interferometry that uses a hybrid of counter-propagating Raman pulses and co-propagating Raman pulses [40, 41]. Of course, the co-propagating Raman pulses can be substituted with microwave pulses. For concreteness, we assume that only the counter-propagating Raman pulses and the microwave pulses would be used. In this paper, we describe how the SCSP and the GESP protocol can be realized using this version of atom interferometry. In addition, we show how to augment these protocols to accommodate large momentum transfer (LMT).

It is not obvious whether a dual-species atom interferometer can adopt the technique of combining spin squeezing and LMT. One immediate problem of incorporating cavity-assisted OATS is that the two species must be squeezed individually to prevent entanglement between the two species. In this paper, we also propose a scheme for a dual-species atom interferometer that employs both spin squeezing and LMT.

The rest of the paper is organized as follows. In Sec. 2, we illustrate the experimental scheme and the sensitivity analysis for atom interferometer employing spin squeezing for acceleration sensing. In Sec. 3, we discuss how to adapt the spin-squeezed atom interferometer to a dual-species atom interferometer for the EP test. The conclusion is given in Sec. 4.

## 2. Spin-squeezed atom interferometer for acceleration sensing

In order to describe the LMT augmented and spin-squeezed atom interferometry protocols, we first introduce the notations of the relevant states. Each atom is modeled as a three-level system with the two ground states denoted as $|\pm\hat{\mathbf{z}}_0\rangle$, and the excited state denoted as $|e\rangle$. In practice, the two ground states are typically the $m_F = 0$ Zeeman substates of the two hyperfine ground states of an alkali atom. The microwave will couple these two ground states. The Raman beams coupling the states $|\pm\hat{\mathbf{z}}_0\rangle$ and $|e\rangle$ is denoted as $\mathbf{k}_\pm$. This three-level system can be reduced to a two-level system, with the effective wavenumber of the pair of counter-propagating Raman beams expressed as $\mathbf{k}_{\text{eff}} = (\mathbf{k}_+ - \mathbf{k}_-)$ if the direction of $\mathbf{k}_{\text{eff}}$ is defined to be that of $\mathbf{k}_+$. The absolute value of the effective wavenumber can be expressed as $k_{\text{eff}} = (k_+ + k_-) \approx 2k_\pm$, where $k_\pm$ is the absolute value of the corresponding wavenumber.

The two-level system can be modeled as spin-1/2 pseudo-spinors, with the spin operator defined as $\mathbf{s} = (s_x, s_y, s_z)$. In this notation, the subscripts $\{x, y, z\}$ represent the three dimensions of the Bloch sphere rather than dimensions of the physical space.

The five OATS-based protocols mentioned in the Introduction, namely SCSP-e, SCSP-o, GESP-e, GESP-o, and the CESP, all involve very similar sequences of pulses [42], and can be denoted as variations of the echo squeezing protocol (ESP). The differences are only the rotation axes of microwave pulses and the duration of the squeezing and unsqueezing pulses. Here, we describe specifically the SCSP-e. Later on, we will discuss how the protocol can be modified easily to realize the GESP or CESP protocols. Figure 1(a) shows the basic version of the pulse sequence of the echo squeezed atom interferometer where the momentum transfer is $2\hbar k_{eff}$ (in contrast to the conventional atom interferometer where the momentum transfer is $\hbar k_{eff}$). In this protocol, all microwave pulses cause rotations around the $y$-axis. The pulse sequence of the protocol is as follows. The atoms are initially in state $|+\hat{\mathbf{z}},0\rangle$, which is a collective state where all the atoms are in state $|+\hat{\mathbf{z}}_0\rangle$ with a linear momentum of zero along the propagation direction of the Raman beams. Application of the first microwave $\pi/2$ pulse is followed by the OATS process with $\mu = \pi/2$, which produces the so-called $x$-directed Schrödinger cat state, defined as $(|+\hat{\mathbf{x}},0\rangle + e^{i\varphi}|-\hat{\mathbf{x}},0\rangle)/\sqrt{2}$. Here, $\varphi$ is a phase resulting from the squeezing process. The OATS operation is followed by the second microwave $\pi/2$ pulse, which creates a $z$-directed Schrödinger cat state $(|-\hat{\mathbf{z}},0\rangle + e^{i(\varphi+\varphi')}|+\hat{\mathbf{z}},0\rangle)/\sqrt{2}$, where $\varphi'$ is a phase resulting from this auxiliary rotation. It should be noted that the phase $\varphi + \varphi'$ is irrelevant and will be canceled during the inverse of the auxiliary rotation and the squeezing operation. As such, for simplicity, we assume it to be zero. The first counter-propagating Raman $\pi$ pulse ($A_1$) transforms $|\pm\hat{\mathbf{z}},0\rangle$ to $|\mp\hat{\mathbf{z}},\pm\hbar k_{eff}\rangle$, resulting in beam-splitting. Next, another Raman $\pi$ pulse ($B_{-1}$) transforms $|\mp\hat{\mathbf{z}},\pm\hbar k_{eff}\rangle$ to $|\pm\hat{\mathbf{z}},0\rangle$. This is followed by the application of a microwave $\pi$ pulse, and then another Raman $\pi$ pulse ($B_1$). After the application of the final Raman $\pi$ pulse ($C_{-1}$), the state of the atoms can be expressed as $(|+\hat{\mathbf{z}},0\rangle + e^{iN\psi}|-\hat{\mathbf{z}},0\rangle)/\sqrt{2}$, where $\psi$ is the acceleration-induced phase shift for a single atom. Pulse $C_{-1}$ is followed by a $\pi/2$ microwave pulse that transforms the state to $(|+\hat{\mathbf{x}},0\rangle + e^{iN\psi}|-\hat{\mathbf{x}},0\rangle)/\sqrt{2}$. Next, the application of the inverse of the OATS operation induces the interference between the upper arm and the lower arm, which produce the state $\cos(N\psi/2)|+\hat{\mathbf{x}},0\rangle + \sin(N\psi/2)|-\hat{\mathbf{x}},0\rangle$. The final microwave $\pi/2$ pulse rotates the state to $\cos(N\psi/2)|-\hat{\mathbf{z}},0\rangle + \sin(N\psi/2)|+\hat{\mathbf{z}},0\rangle$ so that the phase shift can be measured by detecting the population of state $|\pm\mathbf{z}_0\rangle$. We can see that the phase shift is magnified by a factor of $N$ in this case. This is why the protocol stated above can enhance the sensitivity of the an atom interferometer. Figure 1(b) shows the pulse sequences for the atom interferometer with a momentum transfer of $4\hbar k_{\text{eff}}$. The protocol shown in Figure 1(b) contains four additional Raman $\pi$ pulse $A_2$, $B_{-2}$, $B_2$, $C_{-2}$. Pulse $A_2$ transform $|\mp\hat{\mathbf{z}},\pm\hbar k_{\text{eff}}\rangle$ to $|\pm\hat{\mathbf{z}},\pm 2\hbar k_{\text{eff}}\rangle$, and thus increase the momentum transfer. The rest of the additional pulses are necessary for converging the two arms since they are separated farther away. This approach can be extended to produce even larger momentum transfer. For example, in Figure 1(b), we illustrate the case where the momentum transfer is $6\hbar k_{\text{eff}}$.

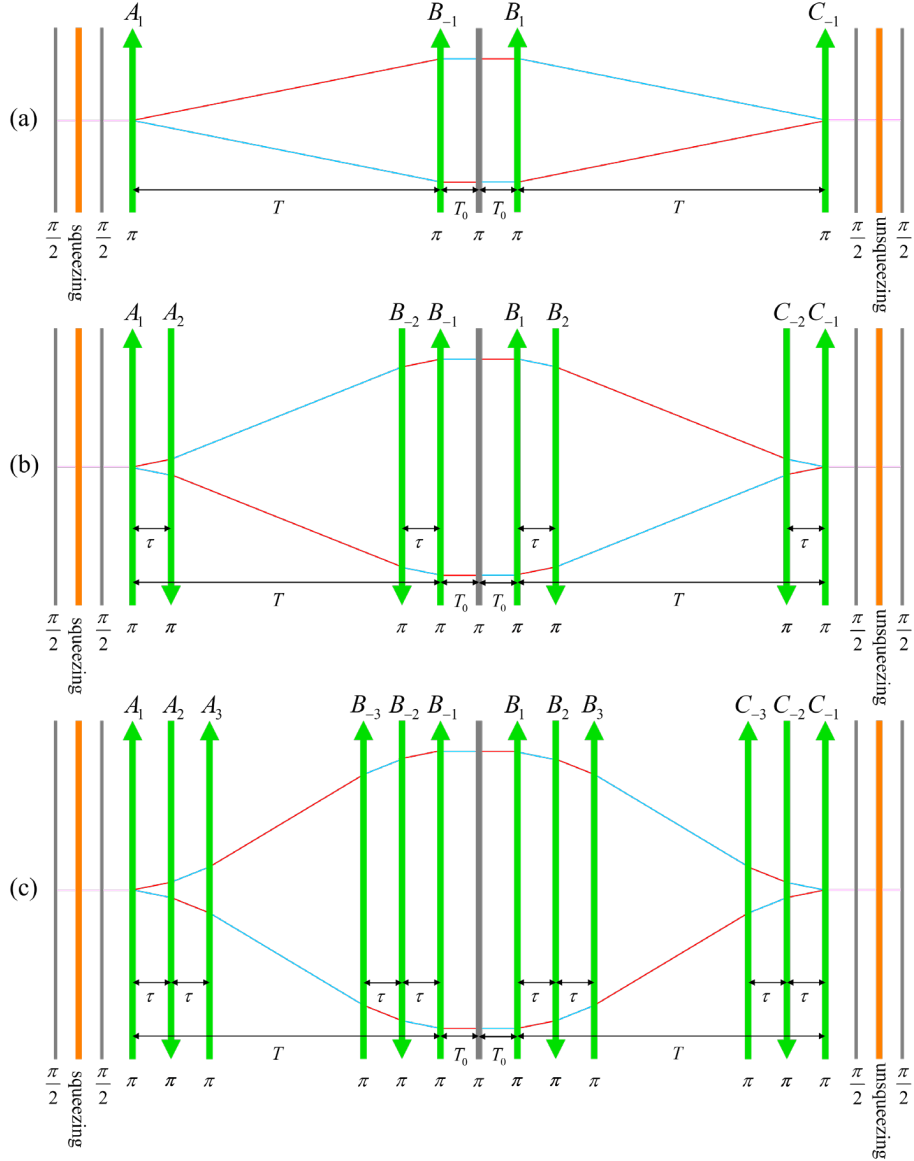

Figure 1: Pulse sequences of (a) the basic version of the spin squeezing protocol for the atom interferometer with a momentum transfer of the spin squeezing protocol for the atom interferometer with a momentum transfer of $2\hbar k_{eff}$, (b) the spin squeezing protocol for the atom interferometer with a momentum transfer of $4\hbar k_{eff}$, (c) the spin squeezing protocol for the atom interferometer with a momentum transfer of $6\hbar k_{eff}$. The green arrows represent the Raman pulses, and the gray lines represent the microwave pulses. The blue (red) lines represent states $|\pm\hat{\mathbf{z}}\rangle$.

The other protocols can be viewed as modifications of the SCSP-e. In SCSP-o, all the microwave pulses except the first and the last ones are around the $x$-axis. The steps in the GESP-e(o) are the same as those for SCSP-e(o), except that the squeezing and un-squeezing interaction times are shorter, corresponding to $\mu < \pi/2$. Thus, GESP-e(o) becomes identical to the SCSP-e(o) for $\mu = \pi/2$. The CESP differs from the GESP-o only in the very last step, where the the last microwave $\pi/2$ pulse causes a rotation around the $x$ axis.

We next calculate the value of $\psi$, the acceleration-induced phase shift for a single atom. The propagator for a microwave $\pi/2$ pulse representing a rotation around the $x(y)$-axis can then be expressed as $\exp(-i(\pi/2)s_{x(y)})$. The effective Hamiltonian of the Raman pulse, which couples states differing in linear momenta by $\hbar k_{\text{eff}}$, can be expressed, in the absence of two-photon detuning, as

$$H = \frac{1}{2}\begin{bmatrix} 0 & -i\Omega e^{-i\phi} \\ i\Omega e^{i\phi} & 0 \end{bmatrix} \tag{1}$$

where $\phi = \mathbf{k}_{\text{eff}} \cdot \mathbf{r}$ is the phase of the light field at the position of the atom. Of course, the value of $\phi$ must be adjusted to account for any overall phase difference between Raman beat signal and the microwave field. However, it can be easily checked that this phase shift does not play a role in these protocols. The corresponding propagators for the Raman $\pi$ pulses can be expressed as:

$$U_\pi = \begin{bmatrix} 0 & -e^{-i\phi} \\ e^{i\phi} & 0 \end{bmatrix} = e^{-i2\phi s_z}e^{-i\pi s_y} = e^{-i\pi s_y}e^{i2\phi s_z} \tag{2}$$

where $\phi = 0$ corresponds to the pulses that represent rotations around the $y$ axis. It should be noted that these propagators only contains the linear terms of the spin operators in the exponent, so that the corresponding propagators for the ensemble can be obtained simply by replacing the single atom operators $s_w$ ($w = x, y, z$) with the collective spin operators $S_w$.

For concreteness, we first focus on the GESP-e, in which all the pulses are around the $y$ axis. With the propagator of a $\pi$ pulse shown in Eq. (2), the net propagator for pulse $A_1$ and $B_{-1}$ can be calculated to be

$$\left(e^{-i2(\phi_{A_1}+\phi_0)S_z}e^{-i\pi S_x}\right)\left(e^{-i\pi S_x}e^{i2(\phi_{B_{-1}}+\phi_0)S_z}\right)=\pm e^{i2(\phi_{B_{-1}}-\phi_{A_1})S_z}=\pm e^{i2\mathbf{k}_{\text{eff}}\cdot(\mathbf{r}_{B_{-1}}-\mathbf{r}_{A_1})S_z} \quad (3)$$

where the sign $\pm$ is evaluated to be $+(-)$ if the number of atoms is even (odd). However, this sign issue is irrelevant in this context. It can be seen from Eq. (3) that two consecutive $\pi$ pulses is equivalent to a rotation around the $z$ axis on the Bloch sphere, with the rotation angle proportional to the displacement of the atoms between $A_1$ and $B_{-1}$. Similarly, the net propagator for pulse $B_1$ and $C_{-1}$ is calculated to be $\pm e^{i2\mathbf{k}_{\text{eff}}\cdot(\mathbf{r}_{C_{-1}}-\mathbf{r}_{B_1})S_z}$. Therefore, the equivalent propagator for the pulses between the squeezing and the unsqueezing operations can be calculated to be

$$\begin{aligned}
&e^{-i(\pi/2)S_y}e^{i2\mathbf{k}_{\text{eff}}\cdot(\mathbf{r}_{C_{-1}}-\mathbf{r}_{B_1})S_z}e^{-i\pi S_y}e^{i2\mathbf{k}_{\text{eff}}\cdot(\mathbf{r}_{B_{-1}}-\mathbf{r}_{A_1})S_z}e^{-i(\pi/2)S_y} \\
&=e^{-i(\pi/2)S_y}e^{i2\mathbf{k}_{\text{eff}}\cdot(\mathbf{r}_{C_{-1}}-\mathbf{r}_{B_1})S_z}e^{-i2\mathbf{k}_{\text{eff}}\cdot(\mathbf{r}_{B_{-1}}-\mathbf{r}_{A_1})S_z}e^{-i\pi S_y}e^{-i(\pi/2)S_y} \\
&=e^{-i(\pi/2)S_y}e^{i2\mathbf{k}_{\text{eff}}\cdot(\mathbf{r}_{A_1}-\mathbf{r}_{B_{-1}}-\mathbf{r}_{B_1}+\mathbf{r}_{C_{-1}})S_z}e^{-i(3\pi/2)S_y} \\
&=\pm e^{-i2\mathbf{k}_{\text{eff}}\cdot(\mathbf{r}_{A_1}-\mathbf{r}_{B_{-1}}-\mathbf{r}_{B_1}+\mathbf{r}_{C_{-1}})S_x}
\end{aligned} \quad (4)$$

which is a rotation around the $x$ axis on the Bloch sphere. Here, we have assumed that there is no Sagnac effect induced phase shift due to rotation. We can see that the net effect of the pulses between the squeezing and the unsqueezing process is only a rotation around the $x$ axis on the Bloch sphere by an angle of the phase shift. Therefore, it can be seen that the protocol shown in Figure 1(a) is equivalent to the fundamental three-step GESP-e described in Ref. [42], with the phase shift $\psi=2\mathbf{k}_{\text{eff}}\cdot(\mathbf{r}_{A_1}-\mathbf{r}_{B_{-1}}-\mathbf{r}_{B_1}+\mathbf{r}_{C_{-1}})$, which is calculated to be $2\mathbf{k}_{\text{eff}}\cdot\mathbf{a}(T^2+2TT_0)$, where the time intervals $T$ and $T_0$ are as defined in Figure 1(a). Accordingly, this basic version of GESP-e augmented atom interferometer can be used for accelerometry. It should be noted that in the case of $\mu = \pi/2$, for which the GESP-e becomes the same as the SCSP-e, this phase gets magnified by a factor of $N$, as shown earlier. When the concomitant increase of the quantum noise by a factor of $\sqrt{N}$ is taken into account, the sensitivity reaches the Heisenberg limit.

Consider next the atom interferometer augmented with both GESP and LMT, corresponding to the cases shown in Figure 1(b) and (c). In this case, the only difference is that the phase shift becomes $\boldsymbol{k}_{eff} \cdot \left[ \sum_{j=1}^{n} \left( \boldsymbol{r}_{A_j} - \boldsymbol{r}_{B_{-j}} - \boldsymbol{r}_{B_j} + \boldsymbol{r}_{C_{-j}} \right) \right]$, where $n$ is the factor by which the momentum splitting is augmented using the LMT process (e.g. $n = 3$ if the momentum splitting is $6\hbar k_{eff}$, corresponding to the case shown in Figure 1(c)). It can be shown that the phase shift can be expressed as $\psi = 2n\boldsymbol{k}_{eff} \cdot \boldsymbol{a}\left[ T^2 + 2TT_0 - (n-1)T\tau \right]$, where the time interval $\tau$ is as defined in Figure 1(b) and (c). For $T \gg \tau$, the acceleration phase shift can be approximated as $2n\boldsymbol{k}_{eff} \cdot \boldsymbol{a}\left( T^2 + 2TT_0 \right)$. Therefore, the spin-squeezed atom interferometer with LMT magnifies the phase shift by a factor of $2n$ compared to the conventional $\pi/2 - \pi - \pi/2$ interferometer, and thus enhance the sensitivity by a factor of $2n$.

The primary effect of squeezing is to increase the gradient of the signal with respect to the phase shift, $\left| \partial \langle S_z \rangle / \partial \psi \right|$, where $S_z$ is the quantum operator measured. The SCSP can magnify the phase gradient by a factor of $N$ if the parity of $N$ matches the version of the SCSP, but also increase the quantum noise by a factor of $\sqrt{N}$. Therefore, the SCSP can reach the Heisenberg limit for a known parity of $N$. Taking into account the application of LMT, the sensitivity is totally

enhanced by a factor of $2n\sqrt{N}$, compared to the conventional $\pi/2$ - $\pi$ - $\pi/2$ interferometer. Both versions of the GESP work optimally in the interval $4\sqrt{2/N} \leq \mu \leq \pi/2 - \sqrt{2/N}$. In this interval, the phase gradient is magnified by a factor of $N\sin\mu/\sqrt{2}$, and the quantum noise is amplified by factor of $\sqrt{N/2}\sin\mu$. Therefore, the ideal sensitivity is enhanced by a factor of $\sqrt{N/2}$, reaching the Heisenberg limit within a factor of $\sqrt{2}$. Taking into account the application of LMT, the sensitivity is totally enhanced by a factor of $n\sqrt{2N}$, compared to the conventional $\pi/2$ - $\pi$ - $\pi/2$ interferometer. The CESP is optimal only for $\mu = N^{-1/2}$. With this value of $\mu$, the phase gradient is magnified by a factor of $\sqrt{N/e}$ while the quantum noise is not changed. Accordingly, the ideal sensitivity reaches the Heisenberg limit within a factor of $\sqrt{e}$. Again, taking into account the application of LMT, the sensitivity is totally enhanced by a factor of $2n\sqrt{N/e}$, compared to the conventional $\pi/2$ - $\pi$ - $\pi/2$ interferometer. Although ideally, all these ESPs can approach the Heisenberg limit, their actual sensitivities in the absence in the presence of detection noise and decoherence mechanisms can differ a lot. The SCSP is robust against detection noise but vulnerable to decoherence mechanisms, while the CESP is less robust against detection noise but more resistant to decoherence mechanisms. The GESP can balance the properties of these two protocols, and is thus possibly the most promising ESP [42].

## 3. Scheme for testing the equivalence principle

A dual-species atom interferometer can be used to test the equivalence principle [12, 13]. In such an interferometer, the two isotopes are initially captured in the same magneto-optic trap. So far, Bragg pulses have been used in dual-species atom interferometers employing $^{87}$Rb and $^{85}$Rb because a single pair of Bragg beams can address both isotopes. In this way, the effective wavenumber of the Bragg beams, $k_{eff}$, and the half duration of the interferometer sequence, $T$, are naturally the same for both isotopes. Therefore, the acceleration phase shift $k_{eff} a T^2$ only depends on the acceleration $a$ for both isotopes.

In principle, Raman pulses can also be used for such a dual-species atom interferometer. However, two different pairs of monochromatic Raman beams are needed to address the two isotopes. These two pairs of Raman beams can be combined and controlled with the same switch so that they spatially coincide and are exactly synchronized. To make the values of $k_{eff}$ of the two pairs of Raman beams the same, we can produce them from the same laser. To produce the Raman beams for $^{85}$Rb, we can guide a beam from the laser through an EOM (electro-optic modulator) tuned to ~1.5 GHz, which is half the hyperfine splitting of the ground state. In this way, the +1 order and -1 order in the output of the EOM would differ in frequency by the hyperfine splitting and their average value is the laser frequency. Each of these frequency components can be extracted with a Fabry-Perot cavity. Alternatively, two different AOMs (acousto-optic modulators) can be used for generating these frequency components, with one up-shifted and another down-shifted. The pair of counter-propagating Raman beams prepared using either approach would have a $k_{eff}$ that is exactly twice the laser wavenumber. If the Raman beams for $^{87}$Rb are prepared in the same way, their $k_{eff}$ will equal the $k_{eff}$ of the Raman beams for $^{85}$Rb. Of course, it would be necessary to adjust the intensities of the beams to balance the light shifts [50] and make the effective Rabi frequencies for the two isotopes equal.

OATS is realized via non-linear interaction between the atoms and the light in an optical cavity [36, 37, 38, 39]. To ensure that the value of the single-photon Rabi frequency remains as uniform as possible longitudinally, use of a ring cavity is preferred. In the model describing the mechanism of OATS, an atom is considered as a three-level system consisting of two ground states, denoted as $|\pm\hat{\mathbf{z}}_0\rangle$, and an excited state, denoted as $|e\rangle$, as discussed earlier. Ideally, to balance the light shifts of states $|+\hat{\mathbf{z}}_0\rangle$ and $|-\hat{\mathbf{z}}_0\rangle$, the cavity should be tuned to the average frequency of the transitions from $|+\hat{\mathbf{z}}_0\rangle$ to $|-\hat{\mathbf{z}}_0\rangle$ and from $|-\hat{\mathbf{z}}_0\rangle$ to $|e\rangle$ if we assume that the Rabi frequencies of these two transitions are the same. In practice, the states $|+\hat{\mathbf{z}}_0\rangle$ and $|-\hat{\mathbf{z}}_0\rangle$ are, respectively, the $F=3, m_F=0$ ($F=2, m_F=0$) and $F=2, m_F=0$ ($F=1, m_F=0$) Zeeman substates for the case of $^{85}$Rb ($^{87}$Rb), state $|e\rangle$ is the $5P_{3/2}$ manifold, which contains multiple hyperfine states, and the probe beam would be $\sigma^\pm$-polarized. Thus, it is necessary to augment the model to take into account the multiplicity of the hyperfine states in the $5P_{3/2}$ manifolds, and the coupling strengths between $|\pm\hat{\mathbf{z}}_0\rangle$ and the relevant Zeeman substates within the hyperfine states in the $5P_{3/2}$ manifold, in order to determine the optimal cavity resonance for each isotope. This analysis is summarized in Appendix.

If the atoms of both isotopes are present in the cavity simultaneously during the OATS operation, the non-linear interaction would produce entanglement among the atoms from both isotopes. As such, the OATS operation for the two isotopes must be carried out separately with two distinct cavities. In describing the process for the dual species atom interferometry augmented by GESP and LMT, we refer to the steps illustrated earlier in Figure 1. The protocol will start as follows. Atoms for each isotope would be trapped in a separate magneto-optic trap, followed by polarization gradient cooling and evaporative cooling. Each ensemble will then be loaded into a separte dipole force trap (DFT), which would be shifted spatially from each other along the direction of propagation of the Raman pump beams. The first microwave $\pi/2$ pulse, followed by the application of OATS and the auxiliary microwave $\pi/2$ pulse will all be carried out while the atoms are still in the DFTs. After this, the DFTs will be turned off, the two ensembles will me made to overlap by controlling the movement of one isotope, using the process describe next.

Controlling the movement of one isotope can be realized with low-Rabi-frequency Raman transitions, as illustrated schematically in Figure 2. When the dipole traps are turned off, the atoms are in a superposition of state $|+\hat{\mathbf{z}}_0, 0\rangle$ and $|-\hat{\mathbf{z}}_0, 0\rangle$. To give a momentum to, for example, state $|+\hat{\mathbf{z}}_0, 0\rangle$ of $^{85}$Rb, we can apply a species-selective Raman pulse resonant to the transition from $|+\hat{\mathbf{z}}_0, 0\rangle$ to $|-\hat{\mathbf{z}}_0, \hbar k_{\text{eff}}\rangle$ (green arrows in Figure 2). This Raman pulse is detuned from the transition between $|-\hat{\mathbf{z}}_0, 0\rangle$ and $|+\hat{\mathbf{z}}_0, -\hbar k_{\text{eff}}\rangle$ by $\hbar k_{\text{eff}}^2 / m \approx 30$ kHz, where $m$ is the mass of $^{85}$Rb. Therefore, if the effective Rabi frequency of the Raman pulse is much lower than 30 kHz, this Raman pulse will only give state $|+\hat{\mathbf{z}}_0, 0\rangle$ of $^{85}$Rb a momentum kick of $\hbar k_{\text{eff}}$. The same method can be used to impart a momentum kick to the state $|-\hat{\mathbf{z}}_0, 0\rangle$ of $^{85}$Rb in the same direction and amount, by using the another pair of Raman beams (orange arrows in Figure 2). Of course, this method requires that the Doppler broadening caused by thermal expansion is much less than 60 kHz, which is necessary anyway for such an experiment [13]. When the two isotopes overlap, we can stop the $^{85}$Rb atoms with the same method, by applying momentum kicks in the opposite direction. To find the effective Rabi frequency of the Raman pulse that should be used, we have implemented a numerical simulation of the efficiencies of the transition from $|+\hat{\mathbf{z}}_0, 0\rangle$ to $|-\hat{\mathbf{z}}_0, \hbar k_{\text{eff}}\rangle$ and the transition from $|-\hat{\mathbf{z}}_0, 0\rangle$ and $|+\hat{\mathbf{z}}_0, -\hbar k_{\text{eff}}\rangle$. The temperature of the atoms used for the simulation is 10 pK, which is the temperature of atoms in the equivalence-principle-test experiment reported in Ref. [51]. The corresponding Doppler broadening (standard deviation) of

10 pK Rb atoms is ~80 Hz. Using a Blackman pulse envelope [52, 53] instead of a square envelope can also suppress the undesired transition from $|-\hat{\mathbf{z}}_0, 0\rangle$ and $|+\hat{\mathbf{z}}_0, -\hbar k_{eff}\rangle$ due to the absence of the side bumps in the Fourier spectrum. Actually, the power of a Blackman pulse is even more concentrated around the central frequency than a Gaussian pulse. The instant Rabi frequency of a Blackman $\pi$ pulse is defined as $\Omega_{eff}(0.42 + 0.5\cos\Omega_{eff}t + 0.08\cos 2\Omega_{eff}t)/0.84$. The efficiencies of the two transitions as functions of $\Omega_{eff}$ are shown in Figure 3. The red curve shows the efficiency of the desirable transition from $|+\hat{\mathbf{z}}_0, 0\rangle$ to $|-\hat{\mathbf{z}}_0, \hbar k_{eff}\rangle$, and the blue curve the undesirable transition from $|-\hat{\mathbf{z}}_0, 0\rangle$ to $|+\hat{\mathbf{z}}_0, -\hbar k_{eff}\rangle$. We can see that the efficiency of the desired transition is ~1 and the undesirable transition ~0 if $\Omega_{eff}$ is between 1 kHz to 10 kHz. Therefore, this technique to control the movement of one isotope is theoretically feasible. In addition, the technique described in Ref. [12] can suppress the remaining imperfection in the overlap between the two isotopes.

After the anti-auxiliary microwave $\pi/2$ pulse which appears before the unsqueezing pulse, we need to separate the two isotopes. The technique of low-Rabi-frequency Raman transition can still be used for such a purpose. Once the two isotopes are spatially separated, they can be confined in two dipole traps again and undergo the rest of the protocol, namely the unsqueezing process and the last microwave $\pi/2$ pulse.

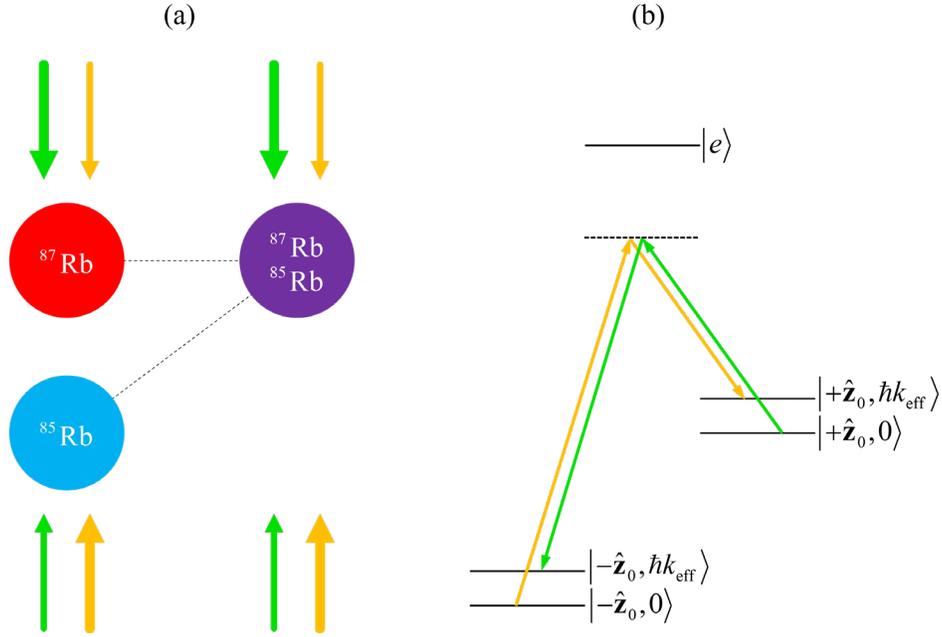

Figure 2 Method to overlap the two isotopes. All the Raman beams only address $^{85}$Rb. The green arrows represent the Raman beams resonant to the transition from $|+\hat{z}_0, 0\rangle$ to $|-\hat{z}_0, \hbar k_{\text{eff}}\rangle$ and the orange arrows the transition from $|-\hat{z}_0, 0\rangle$ to $|+\hat{z}_0, \hbar k_{\text{eff}}\rangle$. The thicker arrows represent the relatively high-frequency Raman beams and the thinner arrows the relatively low-frequency Raman beams. The first Raman pulse give the $^{85}$Rb atoms a momentum of $\hbar k_{\text{eff}}$ upwards. When they overlap the cloud of $^{87}$Rb, the second Raman pulse is applied to stop the $^{85}$Rb atoms.

Practically, it is difficult to make the values of $\mu$ for $^{85}$Rb and for $^{87}$Rb the same. The sensitivity of the spin squeezing protocols depends on the value of $\mu$ and the number of atoms, which will be generally different for these two isotopes. However, a very important aspect of the GESP protocol is that it has essentially the same sensitivity of the Heisenberg limit within a factor of $\sqrt{2}$ for a wide range of $\mu$ [42] and thus is best suited for such a dual-species interferometer.

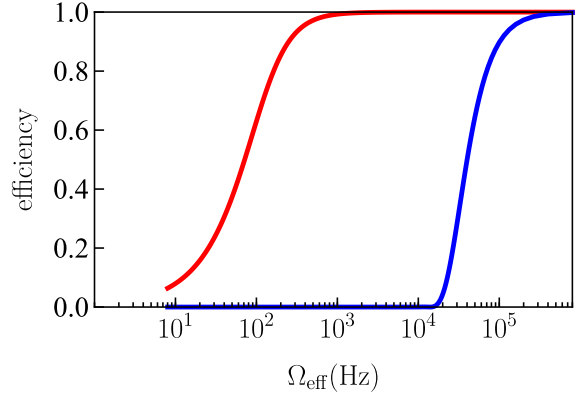

Figure 3 Efficiencies of the two transitions as functions of $\Omega_{\text{eff}}$ driven by a Blackman pulse. The red curve shows efficiency of the desirable transition from $|+\hat{\mathbf{z}}_0, 0\rangle$ to $|-\hat{\mathbf{z}}_0, \hbar k_{\text{eff}}\rangle$, and the blue curve the undesirable transition from $|-\hat{\mathbf{z}}_0, 0\rangle$ to $|+\hat{\mathbf{z}}_0, -\hbar k_{\text{eff}}\rangle$. We can see that the efficiency of the desired transition is ~1 and the undesirable transition ~0 if $\Omega_{\text{eff}}$ is between 1 kHz to 10 kHz.

By modulating the phase of any one of the $\pi$ pulses, denoted as $\phi_{\text{p}}$, we can obtain the signal (namely, the difference between the population of $|+\hat{\mathbf{z}}_0\rangle$ state and the population of $|-\hat{\mathbf{z}}_0\rangle$ state) versus the phase shift. The total phase shift is $2nk_{\text{eff}} aT^2 + \phi_{\text{p}}$. Although the sensitivity can be made the same for these two isotopes, the widths of the fringes as a function of $\phi_{\text{p}}$ depend on the value of $\mu$ and the number of atoms. Assuming, for example, that the value of $\mu$ would be larger for $^{85}$Rb, and the number of atoms trapped for $^{85}$Rb would be higher, the fringes for $^{85}$Rb would be narrower. If the equivalence principle holds, the central peaks, which correspond to the point where $2nk_{\text{eff}} aT^2 + \phi_{\text{p}} = 0$, will coincide, as shown in Figure 4. Therefore, we can lock $\phi_{\text{p}}$ to the central peak (i.e., $\phi_{\text{p}} = -2nk_{\text{eff}} aT^2$) of one isotope and check whether the signal of the other isotope deviates from its central peak.

Ideally, the sensitivity of the GESP can be expressed as $2nk_{\text{eff}} \Delta a T^2 = \sqrt{2}/N$, where $N$ is the number of atoms during the interrogation time. For $n = 5$, $T = 1$ s, and $N = 10^5$, which are close to the values of the parameters adopted in Ref. [13], the uncertainty of the acceleration per shot is calculated to be $8.7 \times 10^{-14}$ m·s$^{-2}$. If the experiment is implemented on the ground or in a low earth orbit, the gravitational acceleration is ~ 9.8 m·s$^{-2}$. Therefore, the relative precision that can be achieved is $\eta = \Delta a/a = 8.8 \times 10^{-15}$ per shot, which is 1600 times the sensitivity reported in Ref. [13]. Of course, the experimental sensitivity reported in Ref. [13] did not reach the standard quantum limit. Compared to the theoretically highest sensitivity of the experiment in Ref. [13], which is the standard quantum limit, the factor of sensitivity enhancement would be $\sqrt{N/2} \approx 220$. It should also be possible to increase the number of atoms by a factor of 10 compared to what was employed in Ref. [13]. In that case, the sensitivity that can be achieved in a single shot would be $\eta = \Delta a/a = 8.8 \times 10^{-16}$, since the sensitivity under the GESP is proportional to $1/N$. If the experiment is implemented on a space-borne platform in a low earth orbit, the value of $T$ can be much larger. If $T = 100$ s, which is reasonable for the temperature of the atoms envisioned above, and $10^5$ shots are implemented (corresponding to approximately 116 days of data collection), the ideally achievable sensitivity would be $\sim 2.2*10^{-22}$.

Of course, in practice it would be exceedingly difficult to reach this ideal limit. As we have discussed at length in Ref. 42, the GESP makes it possible to reach the Heisenberg limit of sensitivity within a factor of $\sqrt{2}$ for a broad range of values of the squeezing parameter. The optimal choice of the squeezing parameter would depend on the nature of the various sources of noise. If the maximal value of the squeezing parameter is used, the effect of all excess detection noise, as well as the noise generated during the squeezing process itself, can be strongly suppressed. On the other hand, in this limit, the effect of collisions with background atoms would become very significant. Use of a cryogenic vacuum [54] may be necessary to suppress the effect of such collisions strongly enough to make the system operate in this regime. Assuming the use of a cryogenic vacuum, and operation of the GESP for the maximal value of the squeezing parameter, while allowing for a factor of ~45 between the ideal and actual values achievable, it may be possible reach a sensitivity of the order of $\sim 10^{-20}$. Of course, the number of shots can be increased to improve the sensitivity further. We would also like to note that the analysis of the effects of various sources of noise as presented in Ref. [42] is not comprehensive, and more detailed studies, theoretical as well experimental, are warranted in order to determine the actual degree of enhancement in sensitivity that could be achieved using the GESP in general, and the dual-species atomic interferometry augmented by LMT and the GESP in particular.

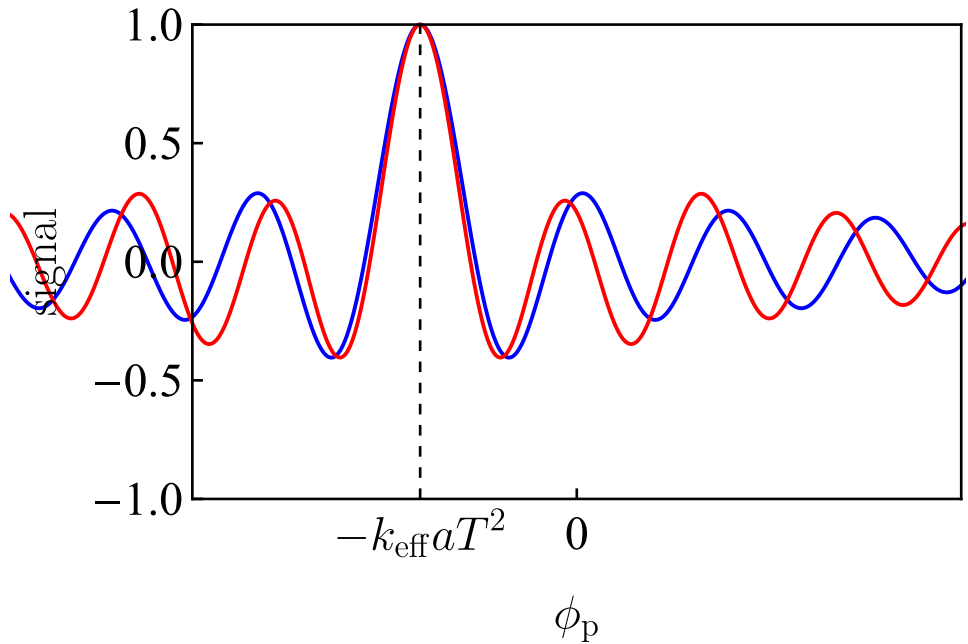

Figure 4 Example of the signals of the generalized echo squeezing protocol for $^{85}$Rb (red) and $^{87}$Rb (blue). The position of the central peak is at $\phi_p = -k_{eff} a T^2$. If the equivalence principle holds, the central peaks of both isotopes will coincide.

Both increasing the interrogation time and increasing the momentum transfer can enhance the sensitivity. However, for an experiment in the weightless environment, one must satisfy the constraint imposed by the length of the vacuum chamber, $x: 2n\hbar k_{eff} T/m \leq x$. Recalling that the phase shift is $\phi = 2nk_{eff} aT^2$, we find that $\phi \leq maxT/\hbar$, which is proportional to $T$ but does not depend on $n$. Therefore, if the physical dimension of the apparatus is the primary constraint, it may be optimal to increase $T$ at the expense of decreasing $n$, while satisfying the constraint $2n\hbar k_{eff} T/m \leq x$. However, other factors may restrict the maximum value of $T$, such as the expansion of the atomic cloud. Taking these factors into account, use of LMT is expected to be useful for increasing the sensitivity of the experiment.

## 4. Conclusion

We have shown theoretically the feasibility of applying spin squeezing to a light pulse atom interferometer even in the presence of large momentum transfer using off-resonant Raman transitions, in order to enhance the sensitivity of accelerometry close to the Heisenberg limit. Even if practical imperfections lower the sensitivity from the ideal level, there is still a good chance to surpass significantly the standard quantum limit with this scheme. We also show how to implement this scheme in a dual-species atom interferometer for precision test of the equivalence principle by measuring the Eötvös parameter. Based on experimental constraints, we find that the generalized echo squeezing protocol, which enhances the sensitivity close to the Heisenberg limit for a very broad range of values of the squeezing parameter, is the best suited for such an experiment. For a space borne platform in low earth orbit, employing a cryogenic vacuum system, such a scheme may enable the measurement of the Eötvös parameter with a sensitivity of the order of $10^{-20}$, using 116 days of data collection.

# Appendix

In this appendix, we calculate the optimal cavity resonance that balances the light shifts of the two ground hyperfine states of Rb. Specifically, we assume that the cavity field would be $\sigma^+$-polarized. We use first $^{87}$Rb as the example to illustrate the calculation, and then show what the result would be for $^{85}$Rb. Consider first the light shift of the ground state $F=2, m_F=0$. The light field inside the cavity couples this state to the $m_F=1$ Zeeman substate of each hyperfine state in the $5P_{3/2}$ manifold. Each coupling contributes a light shift to the ground state $F=2, m_F=0$. Therefore, the total light shift of this ground state is the sum of the light shifts by all the couplings, which can be expressed as

$$\omega_{LS2} = \sum_{j=1}^{3} \frac{|\Omega_{2\to j}|^2}{4(\omega_{HFS}/2 + \omega_{j3} - \Delta\omega)} = \Gamma^2 \sum_{j=1}^{3} \frac{|\alpha_{2\to j}|^2}{4(\omega_{HFS}/2 + \omega_{j3} - \Delta\omega)} \tag{5}$$

where $\Omega_{2\to j}$ is the Rabi frequency of the transition from the ground state $F=2, m_F=0$ to the excited state $F=j, m_F=1$, $\omega_{HFS}$ is the energy difference between the $F=1$ and $F=2$ states, $\omega_{j3}$ is the energy difference between $F=j$ ($j=1,2,3$) and $F=3$ in the $5P_{3/2}$ manifold, $\Delta\omega$ (whose optimal value is to be determined by this analysis) is as defined in Figure 5, $\Gamma$ is the spontaneous decay rate of the $5P_{3/2}$ manifold, and $\alpha_{2\to j}$ is the matrix element for the transition from the ground state $F=2, m_F=0$ to the excited state $F=j, m_F=1$, as shown in Figure 5. In this model, we have assumed that the Rabi frequency of the strongest transition remains much smaller than $\omega_{HFS}/2$. A similar expression is used to determine the total light shift of the ground state $F=1, m_F=0$, denoted as $\omega_{LS1}$. To balance the light shifts, we require that $\omega_{LS1} + \omega_{LS2} = 0$. Solving this equation, we can obtain the optimal value of $\Delta\omega$, which is $2\pi \times 235.7$ MHz for $^{87}$Rb.

It is important to note that this value is independent of the intensity of the probe field, as long the assumption regarding the Rabi frequency of the strongest transition stated above remains valid. Implementing the same calculation, we can also obtain the value of $\Delta\omega$ for $^{85}$Rb, which is $2\pi\times 114.2$ MHz.

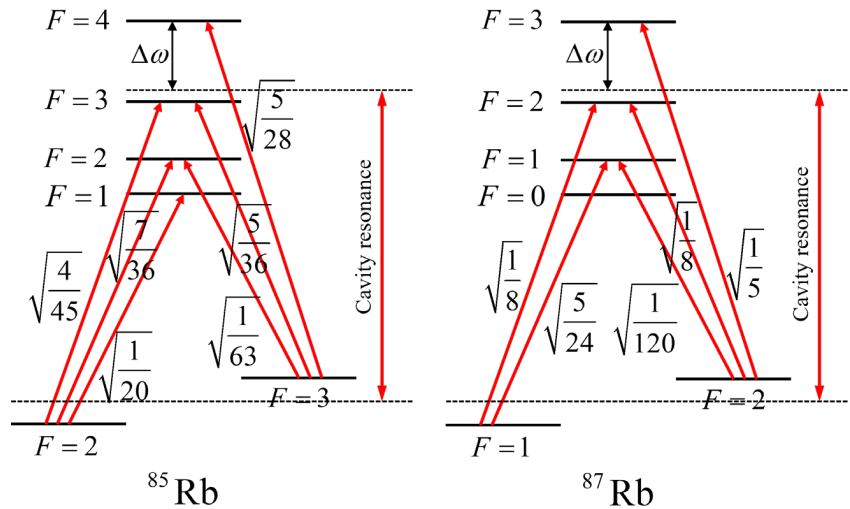

Figure 5 Relevant transitions for calculating the optimal cavity resonance that balance the light shifts of the two ground states.

## Acknowledgement:

This work has been supported equally in parts by NASA grant number 80NSSC20C0161, the Department of Defense Center of Excellence in Advanced Quantum Sensing under Army Research Office grant number W911NF202076, ONR grant number N00014-19-1-2181, and the U.S. Department of Energy, Office of Science, National Quantum Information Science Research Centers, Superconducting Quantum Materials and Systems Center (SQMS) under contract number DE-AC02-07CH11359.

---

[1] NRC Committee on the Physics of the Universe (Chair, M. Turner) report "Connecting Quarks with the Cosmos: Eleven Science Questions for the New Century," 2003, https://nap.nationalacademies.org/login.php?record_id=10079.